\documentclass[onecolumn,aps,pra,reprint,superscriptaddress,longbibliography]{revtex4-1}

\usepackage{amsmath}
\usepackage{graphicx}
\usepackage[lofdepth,lotdepth, caption=false]{subfig}
\usepackage{verbatim}
\usepackage{color}
\usepackage{dcolumn}
\usepackage{bm}
\usepackage{miller}
\usepackage{color}
\usepackage{amsfonts}
\usepackage{setspace}
\usepackage{xr-hyper}
\usepackage{hyperref}
\usepackage[english]{babel}

\usepackage{natbib}
\begin{document}

\title{Exchange anisotropy-driven noncollinear magnetism and magnetic transitions in MnTiO${_3}$ ilmenite}

\author{Srimal Rathnayaka}
\affiliation{Department of Physics, University of Virginia, Charlottesville,
Virginia 22904, USA}

\author{Luke Daemen}
\affiliation{Neutron Scattering Division, Oak Ridge National Laboratory, Oak Ridge, Tennessee 37831, USA}

\author{Despina Louca}
\thanks{Corresponding author}
\email{louca@virginia.edu}
\affiliation{Department of Physics, University of Virginia, Charlottesville,
Virginia 22904, USA}

\begin{abstract}
Evidence for multiple magnetic transitions and unconventional spin exchange interactions in the ilmenite insulator MnTiO${_3}$ is provided via neutron scattering. On cooling, while G-type antiferromagnetic (AFM) order sets in first at 63 K with a k${_1}$ = (000) characteristic wave vector, a weaker second magnetic transition with k${_2}$ = (00$\frac{3}{2}$) appears near 42 K, giving rise to a noncollinear structure. Intrinsic buckling of the honeycomb lattice along $\textit{c}$ creates bond anisotropy and a distorted crystal field that can lead to exchange paths that modulate orbital overlap and spin-orbit coupling. The inelastic spectrum is best described by magnetic exchange anisotropy that breaks the local symmetry of the honeycomb, with competing AFM Heisenberg, Dzyaloshinskii–Moriya and alternate intra-planar ferromagnetic (FM) interactions, that may yield a weakly-coupled ladder system.


\end{abstract}

\maketitle

\begin{figure}[b]
\begin{center}
\includegraphics[width=8cm]{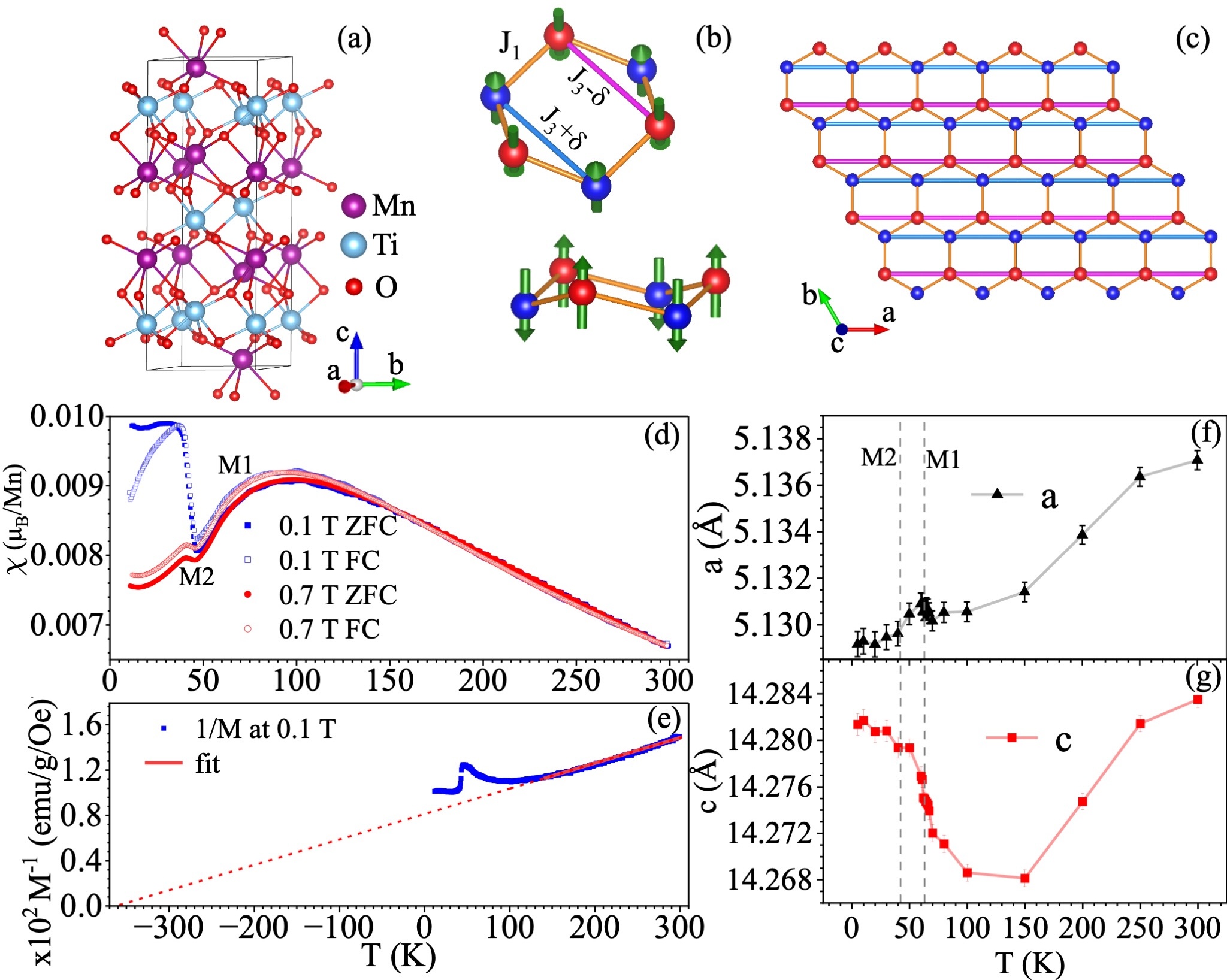}	
\end{center}
\caption{(a) The unit cell of the R$\overline{3}$ ilmenite structure. (b) Top and side view of a honeycomb cell where red and blue indicate Mn sites that are shifted up or down from the plane. J$_1$ and J$_3 \pm \delta$ are in-plane exchange interactions. (c) The bipartite honeycomb network of magnetic Mn sites is bond anisotropic, represented by the different colored lines. (d) The magnetic susceptibility data at 0.1 T and 0.7 T fields for zero-field cooled (ZFC) and field cooled (FC) measurements and (e) its inverse for the 0.1 T measurement. The data were fit using the Curie-Weiss law from 120 to 295 K. (f) The temperature dependence of the lattice constants $a$ and (g) $c$ obtained from the diffraction data refinement. The dotted lines correspond to M1 (G-type) and M2 (A-type) transitions.} 
\label{NewFig1n}
\end{figure}

Honeycomb lattices with AFM order have attracted considerable attention because they can host a wide variety of quantum phenomena ranging from Dirac‑like magnon excitations, Kitaev‑type quantum‑spin‑liquid behaviors \cite{kitaev2006anyons,elliot2021order,li2021identification, yuan2020dirac,pershoguba2018dirac,li2021identification,yuan2020spin,rathnayaka2025magnetic}, to spin-Nernst and thermal Hall effects, driven by broken symmetries and anisotropic exchange \cite{elliot2021order,kitaev2006anyons,rau2014generic,hwan2015direct,songvilay2020kitaev,cheng2016spin,kondo2022nonlinear,kasahara2018majorana}. The ATiO$_3$ (A= Mn, Co, Ni) ilmenites (structure shown in Fig.~\ref{NewFig1n}(a)), add to this growing class of complex materials. Their rhombohedral structure hosts a bipartite honeycomb lattice of magnetic ions exhibiting spin-orbit coupling (SOC) and spin-orbit exciton (SOE) driven magnetic dynamics along with competing exchange interactions as shown in Figs.~\ref{NewFig1n}(b) and ~\ref{NewFig1n}(c)\cite{yuan2020dirac,yuan2020spin,elliot2021order,liu2022spin,rathnayaka2024temperature,kikuchi2025dirac,rathnayaka2025magnetic,hwang2021spin}. Specifically, in MnTiO$_3$, the honeycomb layer is intrinsically buckled (Fig.~\ref{NewFig1n}(b)), with two interpenetrating triangular spin arrangements, antiferromagnetically coupled. This kind of lattice naturally supports geometrically frustrated interactions and unconventional magnetic ground states, making it highly sensitive to anisotropies and competing exchange pathway.

MnTiO${_3}$ exhibits a Neel transition at T$_N$ = 63 K~\cite{stickler1967magnetic} to a G-type structure with spins aligned along the $c$-axis~\cite{goodenough1967theory}. We refer to this phase as M1. This transition is seen as a broad hump in the bulk magnetic susceptibility and the plot shown in Fig.~\ref{NewFig1n}(d) is consistent with previous measurements ~\cite{stickler1967magnetic,maurya2015evidence,mohanty2019neutron,maurya2015temperature,hwang2021spin}. Both in-plane and out-of-plane interactions are antiferromagnetically coupled, consistent with the Goodenough-Kanomori rules in which superexchange interactions through 180$^{\circ}$ bonds favor AFM coupling~\cite{kanamori1959superexchange,goodenough1967theory,osmond1964magnetic}. The magnetic propagation vector in this case is $k_1$ = (000) and the ordered moment is 4.55$\mu_B$/Mn~\cite{shirane1959neutron}. Earlier inelastic neutron scattering measurements reported magnetic excitations up to $\sim$11 meV~\cite{hwang2021spin}. A single magnon band with a maximum energy reaching $\sim$11 meV had been observed and described by a Heisenberg-like magnetic Hamiltonian, with stronger intra-plane and weaker inter-plane AFM exchange interactions. Yet, bulk susceptibility measurements indicate a second magnetic effect below 45 K as shown in Fig.~\ref{NewFig1n}(d) whose origin is unclear~\cite{stickler1967magnetic,maurya2015evidence,hwang2021spin,pal2022competing,mohanty2019neutron,dey2021single,fabritchnyi2003mossbauer}. Studies attributed this feature to possibly a Mn$_3$O$_4$ impurity \cite{dey2021single,fabritchnyi2003mossbauer,gries2022role}. A study on MnTi$_{1-x}$Mn$_x$O$_3$ reported that the anomaly near 42 K becomes more pronounced with doping, for $x \gtrsim 0.07$, possibly due to the increasing presence of Mn$_3$O$_4$~\cite{pal2022competing}, an extrinsic effect. The possibility of this transition being intrinsic to MnTiO$_3$ has also been suggested ~\cite{hwang2021spin}.

 

In this Letter, we show from neutron powder diffraction that the 42 K transition observed in the susceptibility is linked to a second magnetic order. We refer to this phase as M2. Additional magnetic Bragg peaks are present in the diffraction data having the same temperature dependence as the second susceptibility transition. The additional diffraction peaks are associated with a second wave vector $k_2$ = (00$\frac{3}{2}$) and A-type ordering. Combining M1 with M2 yields a spin canted antiferromagnet. From the inelastic neutron scattering spectra, weak magnetic excitations centered around 15 meV are observed, in support of the canted spin structure. Comparison with earlier spin wave measurements on Mn$_3$O$_4$ indicate that its magnetic intensity is not consistent with the magnetic excitations observed here \cite{chung2008magnetic}, excluding the presence of a second phase. No Bragg peaks of Mn$_3$O$_4$ were observed either. 
these findings call for a new theoretical description of MnTiO$_{3}$. Quantitative analysis of the inelastic spectra point to a Hamiltonian with exchange anisotropy that breaks the local symmetry of the honeycomb, competing AFM Heisenberg, Dzyaloshinskii–Moriya due to canting and alternate intra-planar FM interactions due to the bond anisotropy. We argue that our findings show that MnTiO$_{3}$ is a weakly-coupled ladder system. 

The powder sample was prepared using solid-state reaction method. MnO$_2$ and TiO$_2$ powders were mixed in a 1.05:1 molar ratio and reacted at 950$^{\circ}$C under air for 20 hours. Neutron scattering measurements were carried out at the VISION time-of-flight spectrometer of Oak Ridge National Laboratory (ORNL). Diffraction and inelastic data were collected simultaneously between 5 to 300 K. The final neutron energy for inelastic scattering was fixed at 3.5 meV by Bragg-reflection on a series of curved pyrolytic graphite analyzers at nominal scattering angles of 45$^{\circ}$ and 135$^{\circ}$, corresponding to low-Q (LQ) and the high-Q (HQ) paths in (Q,omega)-space with a quadratic dependence of omega on Q in the 1.5 to 30 \AA$^{-1}$ range~\cite{seeger2009resolution}. Refinement of the diffraction data confirmed the presence of a single phase. The dynamic susceptibility and magnon dispersions were simulated using Sunny~\cite{sunny}.

\begin{figure}[t]
\begin{center}
\includegraphics[width=8.4cm]{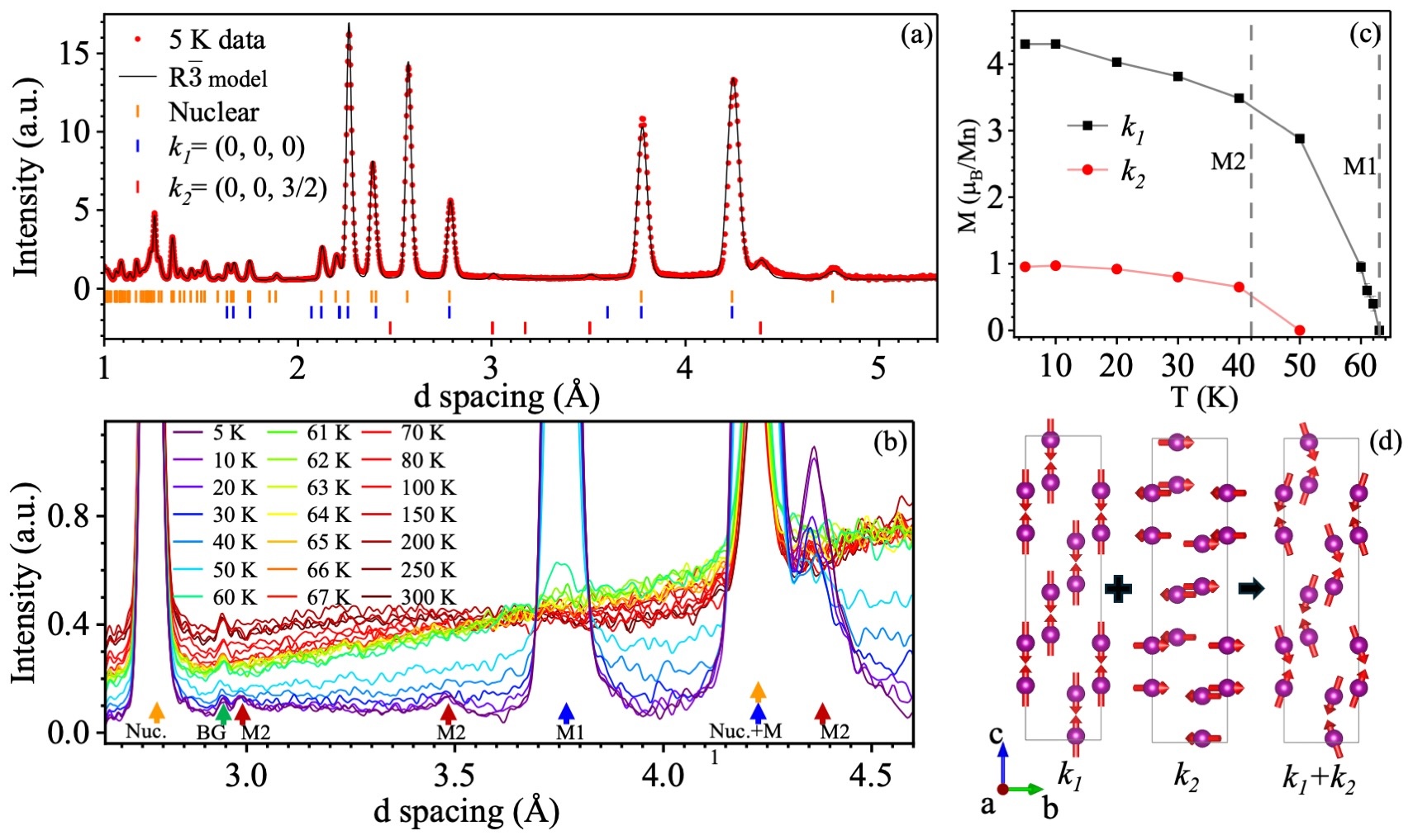}	
\end{center}
\caption{(a) The diffraction data at 5 K is compared to a model calculated using the R$\overline{3}$ symmetry. The orange tick marks correspond to the nuclear Bragg peaks, the blue to the M1 phase with $k_1$ = (000), and red to the M2 phase $k_2$ = (00$\frac{3}{2}$). (b) The temperature dependence of the diffraction pattern in d-spacing, from 2.6 \AA ~to 4.6 \AA. The nuclear peaks are denoted by orange arrows (2.80 \AA). Peaks associated with M1 transition are denoted by blue arrows (3.75 \AA~and 4.25 \AA), and M2 is denoted by red arrows (4.40 \AA,~3.50 \AA,~and 3.00 \AA). The peak at 2.95 \AA ~is a background peak. (c) The order parameters associated with the two magnetic $k$ vectors. The two transitions are linked to distinct propagation vectors, with M1 corresponding to $k_1$, and M2 to $k_2$. Error bars are smaller than data points. (d) The corresponding  $k_1$ and $k_2$ spin structures, and the resultant spin order combining the two $k$ vectors.} 
\label{NewFig2n}
\end{figure}

The magnetic susceptibility, $\chi$, of Fig.~\ref{NewFig1n}(d) shows that the first magnetic transition (M1), occurs around T$_N$ $\sim$ 63 K. This is a broad peak consistent with reported data~\cite{stickler1967magnetic,mohanty2019neutron}. The second transition (M2) is observed near 42 K, where the magnetic susceptibility data show an FM-like transition. 
Shown in Fig.~\ref{NewFig1n}(e) is the inverse $\chi$ using the 0.1 T field-cooled (FC) data. The data were fit using the Curie-Weiss law, $\chi$ = C/(T - $\theta$$_{CW}$)+ $\chi_{0}$. C is the Curie constant and $\theta$$_{CW}$ is the Curie-Weiss temperature. The fit, although limited, yields a $\theta$$_{CW}$ $\approx$ –361 $\pm$ 2 K. Several $\theta$$_{CW}$ values have been reported in the literature ~\cite{mohanty2019neutron, szubka2010electronic,stickler1967magnetic,hwang2021spin}. In Figs.~\ref{NewFig1n}(f) and ~\ref{NewFig1n}(g), the lattice constants $a$ and $c$ obtained from the structure refinements are shown as a function of temperature. The lattice constant $a$ exhibits typical thermal expansion while lattice constant $c$ exhibits negative thermal expansion (NTE) on cooling from $\sim$150 K. These results are in good agreement with earlier reports~\cite{maurya2015evidence,pal2022competing}. The NTE is most prominent with the magnetic transitions, possibly linked to magnetostriction. 

Shown in Fig.~\ref{NewFig2n}(a) is the 5 K neutron diffraction data. The magnetic reflections (indicated in blue tick marks) from the M1 phase are reproduced using the $k_1$ = (000) wave vector~\cite{shirane1959neutron}. The additional weaker magnetic reflections (red tick marks) from the M2 phase cannot be accounted for using the $k_1$ alone or by considering Mn$_3$O$_4$ as the impurity phase. In Fig.~\ref{NewFig2n}(b), the temperature dependence of the neutron powder diffraction from 5 to 300 K is shown. The intensity at 3.75 \AA~and 4.25 \AA~(blue arrows) increases at T$_N$ = 63 K. Additional reflections at 4.40 \AA,~3.50 \AA,~and 3.00 \AA~(red arrows) also exhibit a temperature dependence, but at a lower temperature, starting at $\sim$ 40 K. In contrast, nuclear peaks such as the one at 2.80 \AA~remains essentially temperature independent, while the feature at 2.95 \AA~ is confirmed by background measurements to be a spurion. The peaks marked by blue arrows are consistent with the reported Neel transition (M1) of MnTiO$_3$ and correspond to magnetic Bragg reflections with $k_1$ = (000)~\cite{silverstein2016incommensurate,shirane1959neutron}. The peaks marked by red arrows coincide with the anomaly observed in the magnetic susceptibility near 42 K (M2). The most intense M2 reflection is only $\sim$10$\%$ of the strongest M1 peak and most likely remained undetected in earlier measurements~\cite{silverstein2016incommensurate,shirane1959neutron}.
The temperature dependence of the order parameters is shown in Fig.~\ref{NewFig2n}(c) for the M1 and M2 structures. The $k_1$ component vanishes near $T_N\approx63$ K, while the $k_2$ component appears below 50 K, in agreement with the onset temperatures inferred from the magnetic susceptibility data.

The additional magnetic Bragg peaks are consistent with a second propagation vector, $k_2$ = (00$\frac{3}{2}$). Representational analysis for $k_2$ yields six possible irreducible representations ($\Gamma_1$–$\Gamma_6$). Among these, only $\Gamma_2$, $\Gamma_4$, and $\Gamma_6$ (basis vectors $\psi_2$, $\psi_4$, and $\psi_6$) provide acceptable solutions to the observed $k_2$ reflections. These solutions correspond to structures with AFM alignment within the basal plane. The refined ordered moments at 5 K are $4.302\pm0.043~\mu_B$/Mn for $k_1$ and $0.954\pm0.064~\mu_B$/Mn for $k_2$.

\begin{figure}[t]
\begin{center}
\includegraphics[width=8.3cm]{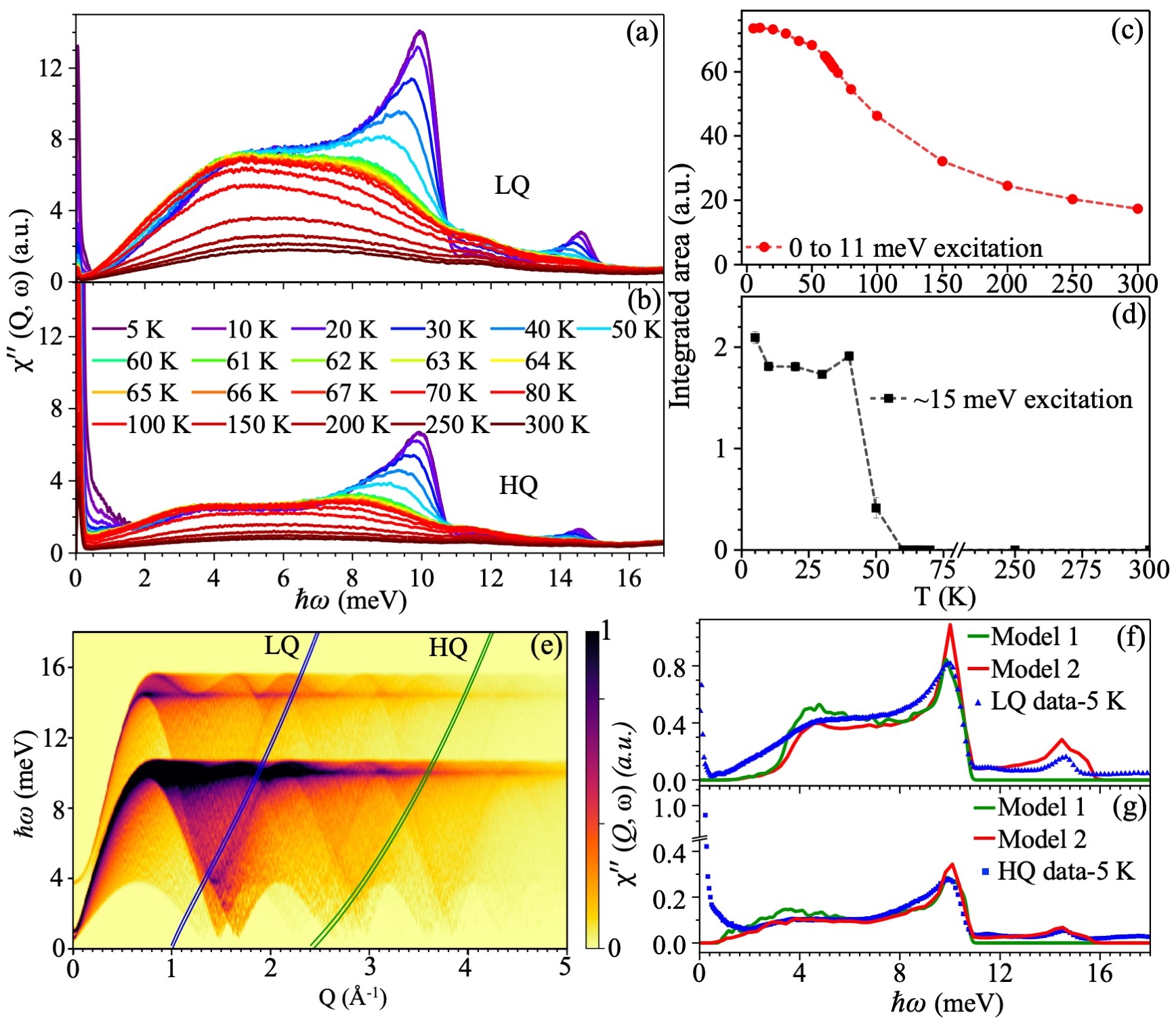}	
\end{center}
\caption{A plot of the dynamic magnetic susceptibility ($\chi^{\prime\prime}$(Q,$\omega$)) vs energy ($\hbar\omega$) as a function of temperature along (a) LQ, and (b) HQ trajectories. Bose correction and background subtraction were performed for the data at all temperatures. (c) The integrated intensity of the LQ spectra from 0 to 12 meV and from 12 to 15 meV in (d). (e) A simulation of the powder averaged spin wave dispersion, S$_\perp$(Q,$\omega$), at 0 K. The exchange constants, J, used for this calculation were, J$_1$=0.70, J$_2$=0.25, J$_3\pm\delta$=-0.32$\pm$0.11, A=-0.008, and D=0.07 meV. The $\chi^{\prime\prime}$(Q,$\omega$) spectrum at 5 K obtained from VISION (blue) along (f) LQ, and (g) HQ paths are compared to a model $\chi^{\prime\prime}$(Q,$\omega$) (green line) determined from a dispersion calculated using J values from Ref.~\cite{hwang2021spin}. A second model $\chi^{\prime\prime}$(Q,$\omega$) (red line) is also shown. This was calculated using J values obtained from this analysis. The calculation accounted for the energy resolution of the instrument and Q resolution set to 0.2 \AA$^{-1}$.} 
\label{NewFig3}
\end{figure}
The $\psi_2$ solution can be excluded, since combining $k_1$ and $k_2$ will produce nonequivalent Mn moments, which is not physical. Also, M1 and M2 contribute independently to the temperature dependence of the magnetic peaks and can be explained by two independent propagation vectors. Therefore, spins must follow the orthogonality condition as we reported in a similar two-magnetic-transition system, NiS$_2$~\cite{yano2016magnetic}. Thus, only $\psi_4$ and $\psi_6$ remain as possible candidates for the $k_2$ structure. The spin arrangement with the $k_1$ component corresponds to the reported G‑type AFM structure, while the $k_2$ component corresponds to A-type with spins lying in the $ab$ plane as shown in Fig.~\ref{NewFig2n}(d) where the exact direction is not for certain as $\psi_4$ has the spin point along $a$, and $\psi_6$ within the $ab$ plane with 23$^\circ$ degrees rotation from $a$ to $b$ axis. The combined $k_1+k_2$ solution yields a canted structure in which spins tilt from the $c$ axis toward the $ab$ plane, with a resultant ordered moment of $4.529\pm0.041~\mu_B$/Mn at 5 K. The moment agrees with the results from earlier studies~\cite{shirane1959neutron}.

\begin{figure}[b]
\begin{center}
\includegraphics[width=8cm]{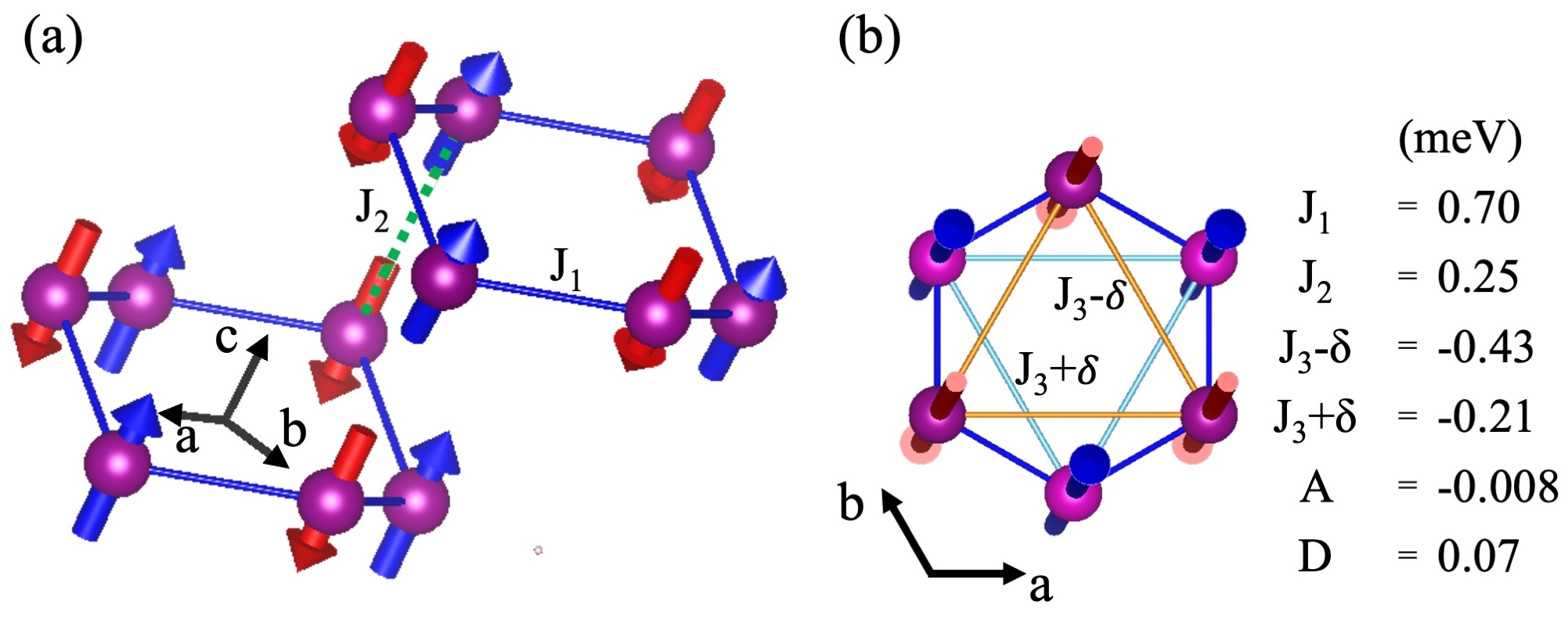}	
\end{center}
\caption{(a) The honeycomb layer with AFM J$_1$ in-plane and AFM J$_2$ out-of-plane. The layers are shifted by $\frac{1}{3}$ along the c-axis. (b) The bi-partite honeycomb cell has two different FM exchange bonds where J$_3$ exchange interactions. The split J$_3$ creates alternate stronger and weaker exchange ferromagnetic paths with $\left( J_{3} \pm \delta \right)$. Also listed are the values used in the Hamiltonian for model 2.} 
\label{NewFig4}
\end{figure}

The dynamic susceptibility, $\chi^{\prime\prime}$(Q,$\omega$), along the LQ and HQ trajectories on VISION are shown in Figs.~\ref{NewFig3}(a) and ~\ref{NewFig3}(b) at multiple temperatures. Two distinct features are observed with a different temperature dependence: a broad, high-intensity feature from 0 to 11 meV, and a low-intensity feature near 15 meV. The first feature can be explained by existing inelastic measurements but not the second. The integrated intensity of the spectra from 0 - 12 meV is shown in Fig.~\ref{NewFig3}(c). With decreasing temperature, the intensity gradually increases with no anomaly at $T_N$, consistent with the magnetic susceptibility data. The persistence of intensity well above $T_N$ reflects the presence of short‑range magnetic correlations~\cite{akimitsu1977magnetic,hwang2021spin}. On the other hand, the integrated intensity around the $\sim$15 meV feature only shows up below 50 K (Fig.~\ref{NewFig3}(d)). In contrast to the low‑energy feature, the intensity of this feature decreases sharply at the second magnetic transition on warming. This behavior supports the origin of the 15 meV excitation to be directly associated with spin canting that develops with the second transition. 

Powder-averaged magnon dispersions were calculated using linear spin wave theory (LSWT) using the Sunny software~\cite{sunny} assuming the G-type AFM structure and the new canted spin structure. The $\chi^{\prime\prime}$(Q,$\omega$) intensities of the LQ and HQ paths from VISION can be compared to intensity integrated along comparable paths in the model dispersions as demonstrated in Fig.~\ref{NewFig3}(e). These comparisons are shown in Figs.~\ref{NewFig3}(f) and ~\ref{NewFig3}(g). The data (in blue) is compared to the two models; model 1 corresponds to the G-type spin structure with a Hamiltonian that consists of an exchange term and a single-ion anisotropy (SIA) term. Only the main feature, up to 11 meV, is reproduced, consistent with Ref.~\cite{hwang2021spin}. In model 2, $\chi^{\prime\prime}$(Q,$\omega$) was calculated using the following Hamiltonian:

\begin{equation}
\begin{aligned}
H &= 
J_{1} \sum_{\langle i,j\rangle}
\mathbf{S}_i \cdot \mathbf{S}_j
 +
\sum_{\langle i,j\rangle}
\left(
    J_{2}\,\mathbf{S}_i \cdot \mathbf{S}_j
    + \mathbf{D}_{ij} \cdot (\mathbf{S}_i \times \mathbf{S}_j)
\right)
\\
&\quad +
\sum_{\langle i,j\rangle}
\left( J_{3} \pm \delta \right)
\mathbf{S}_i \cdot \mathbf{S}_j
\ +
A \sum_i (S_i^{z})^{2},
\end{aligned}
\end{equation}

where S$_i$ is the spin operator on site $i$, and J$_n$ (n= 1, 2, 3) are the first three nearest neighbor exchange interactions. $D_{ij}$ is the Dzyaloshinskii–Moriya interaction (DMI) constant that acts on J$_2$. $A$ is the SIA constant, and $\delta$ is an exchange splitting parameter for J$_3$. Thus, in addition to the exchange and anisotropy terms, this Hamiltonian consists of two additional terms: the DM term, which introduces canting and an alternating FM exchange J$_{3}$ bonds on the honeycomb plane (Figs.~\ref{NewFig1n}(b)). The calculation assumed S = 5/2, the $k_1$+$k_2$ canted magnetic structure, and the magnetic form factor for Mn$^{2+}$. This Hamiltonian yields the magnon dispersions of Fig.~\ref{NewFig3}(e). Cuts across the LQ and HQ paths are plotted as model 2 in Figs.~\ref{NewFig3}(f) and ~\ref{NewFig3}(g). This model reproduces the feature beyond 12 meV. This multi‑exchange Hamiltonian describes the spin dynamics well \cite{ulrich2002magnetic}.

The exchange interactions used in the calculations are illustrated in Figs.~\ref{NewFig4}(a) and \ref{NewFig4}(b), and the corresponding parameters are also listed in the figure. The system is dominated by a strong AFM nearest neighbor interaction, $J_{1}=0.70$~meV, that is significantly larger than the out-of-plane AFM coupling $J_{2}=0.25$~meV. The DMI acts only on the $J_{2}$ bonds, with a DM vector ${D}_{ij}=(D_{ij}^{x},D_{ij}^{y},0)$ of magnitude $D_{ij}^{x}=D_{ij}^{y}=0.07$~meV. This DM term induces a small canting of the spins toward the $ab$ plane, breaking the interlayer collinearity while preserving the collinear arrangement within each layer, as shown in Fig.~\ref{NewFig4}(a). In contrast to the AFM $J_{1}$ and $J_{2}$ interactions, the $J_{3}$ interaction in our model is FM, consistent with the observed spin configuration. The two inequivalent second nearest neighbor in-plane exchange interactions are captured by an exchange-splitting parameter $\delta$, giving $J_{3}\pm\delta=-0.32\pm0.11$~meV, as illustrated in Fig.~\ref{NewFig4}(b). This splitting originates from the strong buckling of the honeycomb layers, which modifies the trigonal-prismatic crystal field and lifts the degeneracy of the $d_{xz}$ and $d_{yz}$ orbitals, thereby generating pronounced bond anisotropy between the $+z$ and $-z$ shifted Mn sites from the ideal honeycomb lattice position. Splitting of J$_3$ lifts the degeneracy of the dispersion bands and results in the 15 meV mode.


While this model that splits J$_3$ captures the 15 meV excitation, the calculated intensity of this feature is not in perfect agreement with the intensity observed experimentally, as shown in Figs.~\ref{NewFig3}(f) and ~\ref{NewFig3}(g). In the vanadate perovskites YVO$_3$ and LuVO$_3$, where spin canting and orbital Peierls effect are present, the magnon spectral weight is likely influenced by orbital effects and requires calculations beyond LSWT~\cite{ulrich2002magnetic,skoulatos2015jahn}. Therefore, the differences might be due to several factors, such as unaccounted magnon-magnon interactions at low temperatures, finite‑temperature spin fluctuations, bond-dependent anisotropies associated with the alternating‑$J_3$ network, and orbital effects~\cite{garcia2025magnon,dyson1956thermodynamic,chen2022anisotropic,rathnayaka2024temperature,rathnayaka2025magnetic,ulrich2002magnetic,skoulatos2015jahn}. 


In the honeycomb lattice of MnTiO$_{3}$, the out‑of‑plane buckling breaks the equivalence of the local honeybomb cell and changes the crystal field for the d$_{xz}$ and d$_{yz}$ orbitals. As a result, the exchange interactions become bond‑anisotropic, with each bond direction favoring different orbital overlap pathways and thus different magnetic couplings. This allows for anisotropic exchange, off‑diagonal terms, and DM interactions on bonds where inversion symmetry is broken. The bond anisotropy can generate dimerization where, along one direction, spins pair into dimers and the system behaves like a weakly coupled chain or a ladder system. Such a system can host AFM exchange, FM exchange and anisotropic SOC driven terms such as DM and competition can drive non-collinear magnetic order, or rung‑like dimerization. These bond‑selective interactions resemble the ones observed in $\alpha$-RuCl$_{3}$ and Na$_{2}$IrO$_{3}$~\cite{banerjee2016proximate,choi2012spin}.

\section*{Data Availability}
\label{Data Availability}

The data that support the findings of this article are openly available~\cite{mto2025github}.


The work at the University of Virginia was supported by the Department of Energy, Grant number DE-FG02-01ER45927. This research used resources at the Spallation Neutron Source, a DOE Office of Science User Facilities operated by the Oak Ridge National Laboratory. The beam time allocation on VISION was through proposal number IPTS-33303.1.



\bibliography{bibliography}

@article{yuan2020dirac,
  title={Dirac magnons in a honeycomb lattice quantum {XY} magnet {CoTiO$_3$}},
  author={Yuan, B. and Khait, I. and Shu, G. J. and Chou, F. C. and Stone, M. B. and Clancy, J. P. and Paramekanti, A. and Kim, Y. J.},
  journal={Physical Review X},
  volume={10},
  number={1},
  pages={011062},
  year={2020},
  publisher={APS}
}

@article{pershoguba2018dirac,
  title={Dirac magnons in honeycomb ferromagnets},
  author={Pershoguba, S. S. and Banerjee, S. and Lashley, J. C. and Park, J. and {\AA}gren, H. and Aeppli, G. and Balatsky, A. V.},
  journal={Physical Review X},
  volume={8},
  number={1},
  pages={011010},
  year={2018},
  publisher={APS}
}

@article{li2021identification,
  title={Identification of magnetic interactions and high-field quantum spin liquid in $\alpha$-{RuCl$_3$}},
  author={Li, H. and Zhang, H. K. and Wang, J. and Wu, H. Q. and Gao, Y. and Qu, D. W. and Liu, Z. X. and Gong, S. S. and Li, W.},
  journal={Nature Communications},
  volume={12},
  number={1},
  pages={4007},
  year={2021},
  publisher={Nature Publishing Group UK London}
}

@article{yuan2020spin,
  title={Spin-orbit exciton in a honeycomb lattice magnet {CoTiO}$_3$: {R}evealing a link between magnetism in d-and f-electron systems},
  author={Yuan, B. and Stone, M. B. and Shu, G. J. and Chou, F. C. and Rao, X. and Clancy, J. P. and Kim, Y. J.},
  journal={Physical Review B},
  volume={102},
  number={13},
  pages={134404},
  year={2020},
  publisher={APS}
}

@article{kitaev2006anyons,
  title={Anyons in an exactly solved model and beyond},
  author={Kitaev, A.},
  journal={Annals of Physics},
  volume={321},
  number={1},
  pages={2},
  year={2006},
  publisher={Elsevier}
}

@article{elliot2021order,
  title={Order-by-disorder from bond-dependent exchange and intensity signature of nodal quasiparticles in a honeycomb cobaltate},
  author={Elliot, M. and McClarty, P. A. and Prabhakaran, D. and Johnson, R. D. and Walker, H. C. and Manuel, P. and Coldea, R.},
  journal={Nature Communications},
  volume={12},
  number={1},
  pages={3936},
  year={2021},
  publisher={Nature Publishing Group UK London}
}

@article{cheng2016spin,
  title={Spin Nernst effect of magnons in collinear antiferromagnets},
  author={Cheng, R. and Okamoto, S. and Xiao, D.},
  journal={Physical Review Letters},
  volume={117},
  number={21},
  pages={217202},
  year={2016},
  publisher={APS}
}

@article{liu2022spin,
  title={Spin waves in layered antiferromagnets with honeycomb structure},
  author={Liu, A. and Finkel'Stein, A. M.},
  journal={Physical Review B},
  volume={105},
  number={21},
  pages={214409},
  year={2022},
  publisher={APS}
}

@article{rathnayaka2024temperature,
  title={Temperature dependence of magnetic excitations in the topological insulator {CoTiO}$_3$},
  author={Rathnayaka, S. and Daemen, L. and Schneeloch, J. A. and Cheng, Y. and Louca, D.},
  journal={Physical Review B},
  volume={109},
  number={17},
  pages={174432},
  year={2024},
  publisher={APS}
}

@article{kikuchi2025dirac,
  title={Dirac Magnon in Honeycomb Lattice Magnet {NiTiO}$_3$},
  author={Kikuchi, H. and Ozeki, M. and Kurita, N. and Asai, S. and Williams, T. J. and Hong, T. and Masuda, T.},
  journal={Journal of the Physical Society of Japan},
  volume={94},
  number={2},
  pages={024703},
  year={2025},
  publisher={The Physical Society of Japan}
}

@article{rathnayaka2025magnetic,
  title={Magnetic dynamics in {NiTiO}$_3$ honeycomb antiferromagnet using neutron scattering},
  author={Rathnayaka, S. and Daemen, L. and Hong, T. and Chi, S. and Calder, S. and Schneeloch, J. A. and Cheng, Y. and Li, B. and Louca, D.},
  journal={Physical Review B},
  volume={112},
  number={10},
  pages={104426},
  year={2025},
  publisher={APS}
}

@article{hwang2021spin,
  title={Spin wave excitations in honeycomb antiferromagnet {MnTiO}$_3$},
  author={Hwang, I. Y. and Lee, K. H. and Chung, J. H. and Ikeuchi, K. and Garlea, V. O. and Yamauchi, H. and Akatsu, M. and Shamoto, S.},
  journal={Journal of the Physical Society of Japan},
  volume={90},
  number={6},
  pages={064708},
  year={2021},
  publisher={The Physical Society of Japan}
}

@article{goodenough1967theory,
  title={Theory of the Magnetic Properties of the Ilmenites {MTiO}$_3$},
  author={Goodenough, J. B. and Stickler, J. J.},
  journal={Physical Review},
  volume={164},
  number={2},
  pages={768},
  year={1967},
  publisher={APS}
}

@article{osmond1964magnetic,
  title={Magnetic exchange interactions inilmenites' {MeTiO}$_3$ ({Me= Mn, Fe, Co and Ni})},
  author={Osmond, W. P.},
  journal={British Journal of Applied Physics},
  volume={15},
  number={11},
  pages={1377},
  year={1964},
  publisher={IOP Publishing}
}

@article{kanamori1959superexchange,
  title={Superexchange interaction and symmetry properties of electron orbitals},
  author={Kanamori, J.},
  journal={Journal of Physics and Chemistry of Solids},
  volume={10},
  number={2-3},
  pages={87},
  year={1959},
  publisher={Elsevier}
}

@article{maurya2015evidence,
  title={Evidence of spin lattice coupling in {MnTiO}$_3$: An {x}-ray diffraction study},
  author={Maurya, R. K. and Singh, N. and Pandey, S. K. and Bindu, R.},
  journal={Europhysics Letters},
  volume={110},
  number={2},
  pages={27007},
  year={2015},
  publisher={IOP Publishing}
}

@article{pal2022competing,
  title={Competing magnetic interactions driven tuning of spin-flop field and anomalous thermal expansion in {MnTi$_{1-x}$Mn$_x$O$_3$}},
  author={Pal, P. and Rahaman, A. and Brar, J. and Bindu, R. and Choudhury, D.},
  journal={Journal of Applied Physics},
  volume={132},
  number={18},
  year={2022},
  publisher={AIP Publishing}
}

@article{stickler1967magnetic,
  title={Magnetic resonance and susceptibility of several ilmenite powders},
  author={Stickler, J. J. and Kern, S. and Wold, A. and Heller, G. S.},
  journal={Physical Review},
  volume={164},
  number={2},
  pages={765},
  year={1967},
  publisher={APS}
}

@article{shirane1959neutron,
  title={Neutron diffraction study of antiferromagnetic {MnTiO}$_3$ and {NiTiO}$_3$},
  author={Shirane, G. and Pickart, S. J. and Ishikawa, Y.},
  journal={Journal of the Physical Society of Japan},
  volume={14},
  number={10},
  pages={1352},
  year={1959},
  publisher={The Physical Society of Japan}
}

@article{akimitsu1977magnetic,
  title={Magnetic critical behavior of a quasi two-dimensional antiferromagnet {MnTiO}$_3$},
  author={Akimitsu, J. and Ishikawa, Y.},
  journal={journal of the physical society of japan},
  volume={42},
  number={2},
  pages={462},
  year={1977},
  publisher={The Physical Society of Japan}
}

@inproceedings{mohanty2019neutron,
  title={Neutron diffraction and magnetic behavior of ilmenite {MnTiO}$_3$},
  author={Mohanty, U. and Kaushik, S. D. and Bhatt, H. and Deo, M. N. and Naik, I.},
  booktitle={AIP Conference Proceedings},
  volume={2115},
  number={1},
  pages={030514},
  year={2019},
  organization={AIP Publishing LLC}
}

@article{fabritchnyi2003mossbauer,
  title={M{\"o}ssbauer characterization of tin dopant ions in the antiferromagnetic ilmenite {MnTiO}$_3$},
  author={Fabritchnyi, P. B. and Korolenko, M. V. and Afanasov, M. I. and Danot, M. and Janod, E.},
  journal={Solid state communications},
  volume={125},
  number={6},
  pages={341},
  year={2003},
  publisher={Elsevier}
}

@phdthesis{dey2021single,
  title={Single-crystal growth, Magnetic and Thermodynamic Investigations of Ilmenite Titanates and Lanthanum Nickelates},
  author={Dey, K.},
  year={2021}
}

@inproceedings{szubka2010electronic,
  title={Electronic structure and magnetic properties of {TiO$_2$-MnTiO$_3$} eutectics},
  author={Szubka, M. and Talik, E. and Kolodziejak, K. and Pawlak, D. A.},
  booktitle={Journal of Physics: Conference Series},
  volume={200},
  number={7},
  pages={072097},
  year={2010},
  organization={IOP Publishing}
}

@article{silverstein2016incommensurate,
  title={Incommensurate crystal supercell and polarization flop observed in the magnetoelectric ilmenite {MnTiO}$_3$},
  author={Silverstein, H. J. and Skoropata, E. and Sarte, P. M. and Mauws, C. and Aczel, A. A. and Choi, E. S. and Van Lierop, J. and Wiebe, C. R. and Zhou, H.},
  journal={Physical Review B},
  volume={93},
  number={5},
  pages={054416},
  year={2016},
  publisher={APS}
}

@article{gries2022role,
  title={Role of magnetoelastic coupling and magnetic anisotropy in {MnTiO}$_3$},
  author={Gries, L. and Jonak, M. and Elghandour, A. and Dey, K. and Klingeler, R.},
  journal={Physical Review B},
  volume={106},
  number={17},
  pages={174425},
  year={2022},
  publisher={APS}
}

@article{yano2016magnetic,
  title={Magnetic structure of {NiS$_{2-x}$Se$_x$}},
  author={Yano, S. and Louca, D. and Yang, J. and Chatterjee, U. and Bugaris, D. E. and Chung, D. Y. and Peng, L. and Grayson, M. and Kanatzidis, M. G.},
  journal={Physical Review B},
  volume={93},
  number={2},
  pages={024409},
  year={2016},
  publisher={APS}
}

@article{ulrich2002magnetic,
  title={Magnetic Neutron Scattering Study of {YVO$_3$}: Evidence for an Orbital Peierls State},
  author={Ulrich, C. and Khaliullin, G. and Sirker, J. and Reehuis, M. and Ohl, M. and Miyasaka, S. and Tokura, Y. and Keimer, B.},
  journal={Physical Review Letters},
  volume={91},
  number={25},
  pages={257202},
  year={2003},
  publisher={American Physical Society (APS)}
}

@article{chung2008magnetic,
  title={Magnetic excitations and orbital physics in the ferrimagnetic spinels {MnB$_2$O$_4$(B= Mn, V)}},
  author={Chung, J. H. and Kim, J. H. and Lee, S. H. and Sato, T. J. and Suzuki, T. and Katsumura, M. and Katsufuji, T.},
  journal={Physical Review B—Condensed Matter and Materials Physics},
  volume={77},
  number={5},
  pages={054412},
  year={2008},
  publisher={APS}
}

@misc{mto2025github,
  author       = {Louca, D.},
  title        = {Neutron Scattering Evidence for a Spin-Canted Phase and {P}eierls-Type Exchange
Modulation in {MnTiO$_3$}},
  year         = {2025},
  howpublished = {\url{https://github.com/despinalouca/Neutron-data/tree/main/MnTiO3_data}},
  note         = {GitHub Repository}
}

@article{skoulatos2015jahn,
  title={{Jahn-Teller versus quantum effects in the spin-orbital material LuVO$_3$}},
  author={Skoulatos, M. and Toth, S. and Roessli, B. and Enderle, M. and Habicht, K. and Sheptyakov, D. and Cervellino, A. and Freeman, P. G. and Reehuis, M. and Stunault, A. and others},
  journal={Physical Review B},
  volume={91},
  number={16},
  pages={161104},
  year={2015},
  publisher={APS}
}

@article{rau2014generic,
  title={Generic spin model for the honeycomb iridates beyond the {K}itaev limit},
  author={Rau, J. G. and Lee, E. K. and Kee, H.},
  journal={Physical review letters},
  volume={112},
  number={7},
  pages={077204},
  year={2014},
  publisher={APS}
}

@article{hwan2015direct,
  title={Direct evidence for dominant bond-directional interactions in a honeycomb lattice iridate {Na$_2$IrO$_3$}},
  author={Hwan Chun, S. and Kim, J. and Kim, J. and Zheng, H. and Stoumpos, C. C. and Malliakas, C. D. and Mitchell, J. F. and Mehlawat, K. and Singh, Y. and Choi, Y. and others},
  journal={Nature Physics},
  volume={11},
  number={6},
  pages={462--466},
  year={2015},
  publisher={Nature Publishing Group UK London}
}

@article{songvilay2020kitaev,
  title={Kitaev interactions in the co honeycomb antiferromagnets {Na$_3$Co$_2$SbO$_6$ and Na$_2$Co$_2$TeO$_6$}},
  author={Songvilay, M. and Robert, J. and Petit, S. and Rodriguez-Rivera, J. A. and Ratcliff, W. D. and Damay, F. and Bal{\'e}dent, V. and Jim{\'e}nez-Ruiz, M. and Lejay, P. and Pachoud, E. and others},
  journal={Physical Review B},
  volume={102},
  number={22},
  pages={224429},
  year={2020},
  publisher={APS}
}

@article{kondo2022nonlinear,
  title={Nonlinear magnon spin Nernst effect in antiferromagnets and strain-tunable pure spin current},
  author={Kondo, H. and Akagi, Y.},
  journal={Physical Review Research},
  volume={4},
  number={1},
  pages={013186},
  year={2022},
  publisher={APS}
}

@article{kasahara2018majorana,
  title={Majorana quantization and half-integer thermal quantum Hall effect in a {K}itaev spin liquid},
  author={Kasahara, Y. and Ohnishi, T. and Mizukami, Y. and Tanaka, O. and Ma, S. and Sugii, K. and Kurita, N. and Tanaka, H. and Nasu, J. and Motome, Y. and others},
  journal={Nature},
  volume={559},
  number={7713},
  pages={227--231},
  year={2018},
  publisher={Nature Publishing Group UK London}
}

@article{garcia2025magnon,
  title={Magnon spectrum of altermagnets beyond linear spin wave theory: {M}agnon-magnon interactions via time-dependent matrix product states versus atomistic spin dynamics},
  author={Garcia-Gaitan, F. and Kefayati, A. and Xiao, J. Q. and Nikoli{\'c}, B. K.},
  journal={Physical Review B},
  volume={111},
  number={2},
  pages={L020407},
  year={2025},
  publisher={APS}
}

@article{dyson1956thermodynamic,
  title={Thermodynamic behavior of an ideal ferromagnet},
  author={Dyson, F. J. },
  journal={Physical Review},
  volume={102},
  number={5},
  pages={1230},
  year={1956},
  publisher={APS}
}

@article{chen2022anisotropic,
  title={Anisotropic magnon damping by zero-temperature quantum fluctuations in ferromagnetic {CrGeTe$_3$}},
  author={Chen, L. and Mao, C. and Chung, J. and Stone, M. B. and Kolesnikov, A. I. and Wang, X. and Murai, N. and Gao, B. and Delaire, O. and Dai, P.},
  journal={Nature communications},
  volume={13},
  number={1},
  pages={4037},
  year={2022},
  publisher={Nature Publishing Group UK London}
}

@inproceedings{maurya2015temperature,
  title={Temperature evolution of the crystal structure of {MnTiO$_3$}},
  author={Maurya, R. K. and Singh, N. and Bindu, R. },
  booktitle={AIP Conference Proceedings},
  volume={1665},
  number={1},
  pages={140013},
  year={2015},
  organization={AIP Publishing LLC}
}

@misc{sunny,
	author = {},
	title = {{G}it{H}ub - {S}unny{S}uite/{S}unny.jl: {S}pin dynamics and generalization to {S}{U}({N}) coherent states --- github.com},
	howpublished = {\url{https://github.com/sunnysuite/sunny.jl}},
	year = {},
	note = {},
}

@article{banerjee2016proximate,
  title={Proximate {K}itaev quantum spin liquid behaviour in a honeycomb magnet},
  author={Banerjee, A. and Bridges, C. A. and Yan, J. Q. and Aczel, A. A. and Li, L. and Stone, M. B. and Granroth, G. E. and Lumsden, M. D. and Yiu, Y. and Knolle, J. and others},
  journal={Nature materials},
  volume={15},
  number={7},
  pages={733--740},
  year={2016},
  publisher={Nature Publishing Group UK London}
}

@article{choi2012spin,
  title={Spin waves and revised crystal structure of honeycomb iridate {Na$_2$IrO$_3$}},
  author={Choi, S. K. and Coldea, R. and Kolmogorov, A. N. and Lancaster, T. and Mazin, I. I. and Blundell, S. J. and Radaelli, P. G. and Singh, Y. and Gegenwart, P. and Choi, K. R. and others},
  journal={Physical review letters},
  volume={108},
  number={12},
  pages={127204},
  year={2012},
  publisher={APS}
}

@article{seeger2009resolution,
  title={Resolution of {VISION}, a crystal-analyzer spectrometer},
  author={Seeger, P. A. and Daemen, Luke L. and Larese, J. Z.},
  journal={Nuclear Instruments and Methods in Physics Research Section A: Accelerators, Spectrometers, Detectors and Associated Equipment},
  volume={604},
  number={3},
  pages={719--728},
  year={2009},
  publisher={Elsevier}
}

\end{document}